\documentclass[prl,reprint,amsmath,amssymb,nofootinbib]{revtex4-1}
\usepackage{epsfig}

\def\beq{\begin{equation}}
\def\eeq{\end{equation}}
\def\bea{\begin{eqnarray}} 
\def\eea{\end{eqnarray}}
\def\nn{\nonumber}
\def\gev{\rm GeV}
\def\tev{\rm TeV}

\def\met{\rm MET}

\def\higgs{{\rm Higgs}}
\newcommand{\lsim}{
\mathrel{\hbox{\rlap{\hbox{\lower4pt\hbox{$\sim$}}}\hbox{$<$}}}}
\newcommand{\gsim}{
\mathrel{\hbox{\rlap{\hbox{\lower4pt\hbox{$\sim$}}}\hbox{$>$}}}}

\begin{document}

\title{Fourth Generation Parity}
\author{Hye-Sung Lee$^{1,2,3}$}
\author{Amarjit Soni$^1$}
\affiliation{Department of Physics, Brookhaven National Laboratory, Upton, NY 11973, USA\\
$^2$Department of Physics, College of William and Mary, Williamsburg, Virginia 23187, USA\\
$^3$Theory Center, Jefferson Lab, Newport News, Virginia 23606, USA}
\begin{abstract}
We present a very simple 4th-generation (4G) model with an Abelian gauge interaction under which only the 4G fermions have nonzero charge.
The $U(1)$ gauge symmetry can have a $\mathbb Z_2$ residual discrete symmetry (4G-parity), which can stabilize the lightest 4G particle (L4P).
When the 4G neutrino is the L4P, it would be a neutral and stable particle and the other 4G fermions would decay into the L4P leaving the trace of missing energy plus the standard model fermions.
Because of the new symmetry, the 4G particle creation and decay modes are different from those of the sequential 4G model, and the 4G particles can be appreciably lighter than typical experimental bounds.
\end{abstract}
\maketitle

There are intriguing arguments and considerable interest in the 4th-generation (4G) fermions \cite{Hou:2008xd,Cvetic:2001nr}.
Since we do not have a good understanding of why we have 3 generations in the standard model (SM), it seems sensible to ask  if the 4G also exists.
It is clearly one of the simplest extensions of the SM, whose symmetries do not restrict the number of fermion generations.
Yet there are some issues in the 4G scenario and, in this paper, we will present a new 4G model which can address these issues and predict novel signatures in the Large Hadron Collider (LHC) experiments.

While it is not certain whether the 4G exists or not, one fact is certain: if the 4G exists, there should be an underlying symmetry or mechanism that distinguishes the 4G from the 3 generations of the SM and renders the 4G neutrino enormously more massive than the lighter 3 neutrinos.
This is required by the LEP measurement of the $Z$ width which strongly supports the fact there are only 3 light active neutrinos \cite{PDG}.
This has been a common issue for all the 4G models since the LEP era \cite{King:1992qr}.

We want to emphasize that this symmetry should be considered natural or at least not uncommon to the extent that most new physics models carry an auxiliary symmetry in order to be realistic: $R$-parity for supersymmetry \cite{Farrar:1978xj}, $T$-parity for the Little Higgs \cite{Cheng:2003ju}, $KK$-parity for extra dimension \cite{Georgi:2000ks}, for instance.
While these are all introduced to address phenomenological issues such as proton stability and electroweak (EW) precision tests, they often stabilize a new particle (LSP by $R$-parity, LTP by $T$-parity, LKP by $KK$-parity), which could be a natural dark matter (DM) candidate.
Thus, in a similar fashion, we can envision a natural auxiliary symmetry for the 4G models that could distinguish 4G fermions from the SM fermions so that it can simultaneously (i) explain the LEP $Z$ width measurement, and (ii) provide stability to a new particle that could be a DM candidate.

Another issue of the 4G scenario is that collider bounds on the 4G fermion masses are saturating the maximal values allowed by the perturbative unitarity.
For instance, in a direct search of 4G up-type quark ($u_4$ or $t'$) in a sequential 4G extension of the SM (often called the standard model with 4 generations, SM4) with a decay mode of $t' \to W^+ + b$ assuming $100 \%$ branching ratio, the current bound from the LHC experiments \cite{CMS:2012ab} is close to the perturbative unitarity bound of $\sim 550 ~\gev$ \cite{unitarity}.

In this paper, we introduce a very simple 4G model accompanied by an Abelian gauge symmetry under which only the 4G fermions are charged, thus dubbed as 4G force.
A residual discrete symmetry or {\it 4G-parity} remains after the $U(1)$ symmetry breaking, causing the lightest 4G particle (L4P) stable.
Note that without a gauge origin, a discrete symmetry may be vulnerable to the Planck scale physics \cite{Krauss:1988zc}.

We will discuss how the new symmetry can satisfy the desirable features to be the 4G auxiliary symmetry and consider its implications for the LHC experiments.
We assert that the 4G scenario, when it is accompanied by the auxiliary symmetry, can remain valid even long after the SM4 is excluded by the experiments at the LHC.

Some recent works on 4G models in various contexts can be found in Refs.~\cite{Kong:2010qd,BarShalom:2011zj,FileviezPerez:2011pt,He:2011ti,Borah:2011ve}.

MODEL:
The particle content in our model is the SM particles including 3 right-handed (RH) neutrinos plus an entire 4G fermion multiplet ($Q_4$, $U_4$, $D_4$, $L_4$, $N_4$, $E_4$) as well as a gauge boson $Z'$ of a new $U(1)$ gauge symmetry.
We assume 2 Higgs doublets ($\Phi$ and $\Phi'$) and a Higgs singlet ($S$) whose vacuum expectation value (vev) breaks the $U(1)$.
(We adopt typical notations $Q \equiv (u, d)_L^T$, $U \equiv u_R$, $D \equiv d_R$, etc.)

The $U(1)$ has nonzero charges for the 4G fermions, but zero charges for the SM fermions (see Table~\ref{tab:charges}).
As is well known, the mixture of the $(B-L)$ and $Y$ is the only possible anomaly-free gauge extension of the SM for each generation without introducing additional fermions besides the right-handed neutrino \cite{Weinberg}.
($B$, $L$, $Y$ are Baryon number, Lepton number, Hypercharge, respectively.)
Thus, the $U(1)$ charge
\beq
{\cal Q} = \left[ (B-L) + x\, Y \right]_4
\eeq
is determined uniquely up to the mixing parameter $x$.

Since the SM fermions are not charged under the $U(1)$, one Higgs doublet ($\Phi$) responsible for the SM fermion masses should not carry any $U(1)$ charge either.
The other Higgs doublet ($\Phi'$) that generates the 4G fermion masses should carry nonzero $U(1)$ charge in general to make Yukawa terms gauge invariant.
In the special case of $x=0$, a single Higgs doublet would be enough to couple and give masses to both SM fermions and 4G fermions, but it gets the same constraint as the SM4 Higgs which is incompatible with the recent LHC data.
(See the discussion later.)

The Yukawa terms are given by
\bea
{\cal L} &=& - y_D \bar Q_a \Phi D_b - y_U \bar Q_a \tilde \Phi U_b - \cdots + h.c. \nn \\
         & & - y'_D \bar Q_4 \Phi' D_4 - y'_U \bar Q_4 \tilde \Phi' U_4 - \cdots + h.c. 
\eea
where $a, b = 1 \sim 3$ covering only the SM generations.
($\tilde \Phi \equiv i \tau_2 \Phi^*$ is conjugate of $\Phi$.)
SM fermion masses are given by $m_f = y_f v / \sqrt{2}$ whereas the 4G fermion masses are given by $m_{f_4} = y'_f v' / \sqrt{2}$.
Total vev of $v_\text{EW} = \sqrt{v^2 + v'^2} \simeq 246 ~\gev$ can be maintained, for example, with $v \simeq 225 ~\gev$ and $v' \simeq 100 ~\gev$ or vice versa.
This would limit the 4G fermion masses, but as we will discuss later, such light 4G fermions may be possible in the presence of 4G-parity.

There is a similarity between our model and the well-established Type-I two Higgs doublet model (2HDM) where $\Phi$ couples to all SM fermions and the $\Phi'$ does not couple to any fermions.
(See also Refs.~\cite{Chen:2012wz,Geller:2012wx,Bellantoni:2012ag} for some recent studies on the 2 Higgs doublets with 4G.)

\begin{table}[tb]
\begin{tabular}{|c|c|c||c|l|c|}
\hline
~Field~ & ~$U(1)$ charge~ & ~$\mathbb Z_2$~ & ~Field~ & ~$U(1)$ charge~ & ~$\mathbb Z_2$~ \\
\hline
~$Q_{1 - 3}$~ & $0$       & $0$             & ~$Q_4$~ & ~~~$1/3 + x$  & $1$ \\
~$U_{1 - 3}$~ & $0$       & $0$             & ~$U_4$~ & ~~~$1/3 + 4x$ & $1$ \\
~$D_{1 - 3}$~ & $0$       & $0$             & ~$D_4$~ & ~~~$1/3 - 2x$ & $1$ \\
~$L_{1 - 3}$~ & $0$       & $0$             & ~$L_4$~ & ~~~$-1 - 3x$  & $1$ \\
~$N_{1 - 3}$~ & $0$       & $0$             & ~$N_4$~ & ~~~$-1$            & $1$ \\
~$E_{1 - 3}$~ & $0$       & $0$             & ~$E_4$~ & ~~~$-1 - 6x$  & $1$ \\
\hline
~$\Phi$~         & $0$       & $0$             & \multicolumn{3}{c}{} \\
~$\Phi'$~        & $3x$      & $0$             & \multicolumn{3}{c}{} \\
~$S$~         & $\pm2/3$  & $0$             & \multicolumn{3}{c}{} \\
\cline{1-3}
\end{tabular}
\caption{$U(1)$ charges ${\cal Q} = [(B-L) + x\, Y]_4$ and the residual $\mathbb Z_2$ discrete charges. $0$ $(1)$ means even (odd) under the parity.}
\label{tab:charges}
\end{table}

Discovering $Z'$ would not be straightforward since the SM fermions have zero $U(1)$ charge.
It also allows the $Z'$ to be much lighter than typically assumed TeV scale.
The relevant terms for the $Z'$ mass are given by
\bea
{\cal L} &=& \frac{1}{2} m_{Z_0}^2 Z_0 Z_0 + \Delta^2 Z_0 Z'_0 + \frac{1}{2} m_{Z'_0}^2 Z'_0 Z'_0
\eea
with $m_{Z_0}^2 = \frac{1}{4} g_Z^2 v_\text{EW}^2$, $m_{Z'_0}^2 = g_{Z'}^2 (9 x^2 v'^2 + \frac{4}{9} v_S^2)$, and $\Delta^2 = -\frac{3}{2} g_Z g_{Z'} x v'^2$ .
The $g_Z$ and $g_{Z'}$ are the effective gauge coupling constants for $Z$ and $Z'$, respectively, and $v_S$ is vev of the Higgs singlet $S$.

The $Z$-$Z'$ mixing angle $\xi$ is determined by $\tan 2\xi = 2 \Delta^2 / (m_{Z_0}^2 - m_{Z'_0}^2)$, which is constrained to be tiny by the precise $Z$ pole measurement at LEP ($|\xi| \lsim {\cal O}(10^{-3})$) \cite{Langacker:2008yv}.
The mixing angle is small enough for sufficiently small $g_{Z'} x$ or sufficiently large $g_{Z'} v_S$.
Thus the small mixing can be easily achieved.
We will not consider kinetic mixing through heavy 4G fermion loops between the $U(1)_Y$ and the new $U(1)$ in this paper.
When a Higgs doublet charged under a $U(1)$ is present, interesting phenomenology associated with a tree-level $Z'$-$Z$-Higgs vertex is possible for both heavy and light $Z'$ scenarios \cite{Barger:2009xg,Davoudiasl:2012ag}.

In general, $B-L$ with an arbitrary shift proportional to $Y$ can leave an unbroken residual $\mathbb Z_2$ discrete symmetry $(-1)^{3(B-L)}$, called Matter-parity, under which all matter particles (quarks and leptons) are odd and the others are even as long as the scalar boson whose vev breaks the $U(1)$ has a right charge.\footnote{Matter-parity is equivalent to $R$-parity $(-1)^{3(B-L)+2J}$ in the supersymmetry framework because of the angular momentum conservation.
For discussions on the gauge origin of $R$-parity (or Matter-parity), see Refs.~\cite{Martin:1992mq,Lee:2010hf}.}
This can occur in our $U(1)$ whose charge is basically the same as $[(B-L) + x\, Y]$ except that only the 4G quarks and leptons are charged.
As a result, only the 4G fermions are odd while the other fermions are even under the parity (see Table~\ref{tab:charges}).
\bea
U(1) &\rightarrow& \underline{\mathbb Z_2 ~(\text{4G-parity})} \\
&& {\small \cdot~ \text{4G fermions (odd)}} \nn \\
&& {\small \cdot~ \text{Others (even)}} \nn
\eea
The Higgs doublet $\Phi'$ with nonzero $U(1)$ charge is still even under 4G-parity, and its vev does not break the $\mathbb Z_2$.

\begin{figure}[t]
\begin{center}
\includegraphics[width=0.232\textwidth]{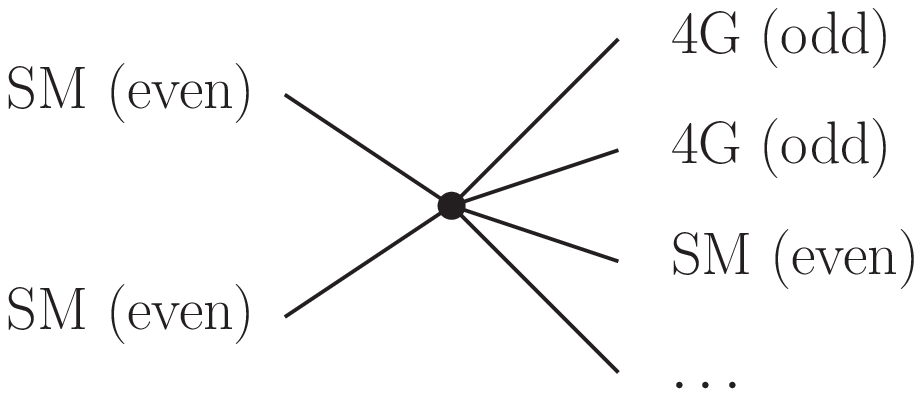} ~
\includegraphics[width=0.232\textwidth]{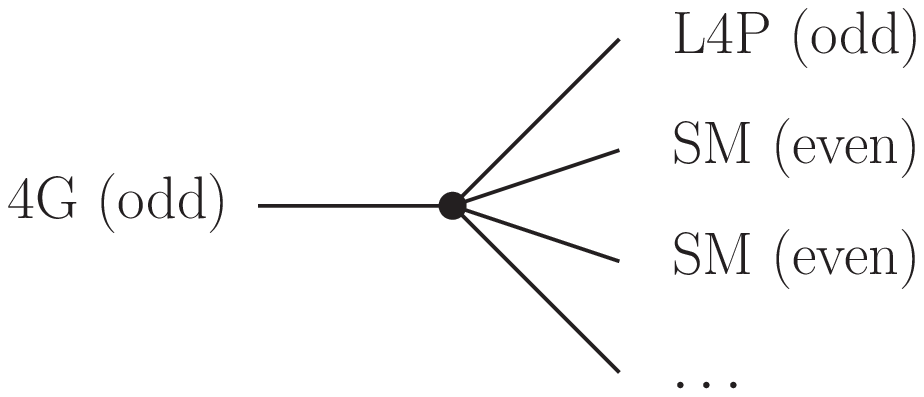}
\end{center}
\caption{Production and decay of 4G fermions in the presence of 4G-parity}
\label{fig:ProductionDecay}
\end{figure}

4G-parity controls the production and decay of the 4G fermions, as illustrated in Fig.~\ref{fig:ProductionDecay}, which is a similar role that $R$-parity plays for superpartners in the supersymmetry models.
It stabilizes the lightest 4G particle or L4P.
In order to avoid a stable charged particle that might conflict with observations, we will take the 4G neutrino as the L4P in this paper.
The 4G-parity also forbids any mixing between the SM fermions and the 4G fermions.
Also severely constrained flavor-changing neutral currents are absent at the tree-level because of the $U(1)$ symmetry.

The $U(1)$ charge of the Higgs singlet $S$ should be $\pm 2$ after normalization of all charges into integers, in order to have a $\mathbb Z_2$ as a remnant discrete symmetry from the $U(1)$  \cite{discrete}.
It is $\pm 2/3$ before the normalization (see Table~\ref{tab:charges}).
Since the 4G RH neutrino ($N_4$) has the $U(1)$ charge $-1$, the renormalizable 4G Majorana mass terms ($m N_4 N_4$ or $S N_4 N_4$) are forbidden by the $U(1)$, and the 4G neutrino is basically a Dirac particle (with only suppressed nonrenormalizable Majorana mass terms).
\bea
{\cal L} &\sim& - y'_N \bar L_4 \tilde \Phi' N_4 + h.c.
\label{eq:4Gnumass}
\eea
Since the seesaw mechanism is irrelevant to this, the 4G neutrino is naturally heavy ($m_{\nu_4} \simeq y'_N v' / \sqrt{2}$) in accord with the LEP $Z$ width measurement.

The $Z$ boson width constrains the 4G fermions including the Dirac $\nu_4$ to be heavier than $m_Z / 2 \simeq 45 ~\gev$.
In addition, the 4G fermions could be produced through the off-shell $Z$ boson at the LEP2 experiment whose highest center-of-mass energy was $207 ~\gev$.
The best lower bound on the 4G fermion mass as large as a half of the LEP2 center-of-mass energy may be obtained except for the stable $\nu_4$.

HIGGS SEARCH:
4G fermions and the Higgs boson have implications for each other in the collider experiments \cite{Kribs:2007nz,Dawson:2010jx}.
While the recent discovery of a new scalar boson of about $125-126 ~\gev$ at the LHC experiments \cite{:2012gk,:2012gu}
needs further study to determine if it is really the long-sought Higgs scalar or not, it would be most natural to take it as a Higgs boson in our model.
The signals are largely consistent with the SM prediction including the $gg \to \higgs \to \gamma\gamma$ mode, which includes the fermions in the loop-induced vertices.
It rules out the 4G models with a single Higgs doublet such as SM4, which predicts Higgs production and decay rates very different from the SM prediction, regardless of the mass of $\nu_4$.
(For detailed discussion about this, see Refs.~\cite{Kuflik:2012ai,Djouadi:2012ae}.)

This situation directs us to limited parameter space that can provide a SM-like Higgs boson, that is, a Higgs scalar that has a cross section and branching ratios similar to the SM Higgs.
Physical Higgs states are in general mixtures of the Higgs scalars after symmetry breakings.
Since the SM-like Higgs should be ``doublet-dominated'' i.e. have couplings to $Z$ and $W$ similar to those of the SM (the singlet composition of a Higgs would not couple to weak gauge bosons because it is not charged under the $SU(2)_L$), let us consider only the doublet mixing here to approximate it.
Assuming the lighter one ($h$) is the SM-like Higgs, its relevant couplings divided by the SM Higgs couplings (for EW gauge boson, SM fermions, 4G fermions, respectively) are given by
$$
r(h VV) = \sin(\beta - \alpha), 
r(h f \bar f) = \frac{\cos\alpha}{\sin\beta}, 
r(h f_4 \bar f_4) = -\frac{\sin\alpha}{\cos\beta}
$$
as one can easily derive using the standard mixing notations in the 2HDM \cite{Branco:2011iw}: $\alpha$ is the mixing angle of 2 neutral Higgs doublets and $\beta$ is defined by $\tan\beta \equiv v / v'$.

The production and decay rate of each mode then can be obtained rather simply by multiplying the coupling ratio to the well-known SM model formula and its straightforward extension for the 4G.
The SM-like Higgs can be obtained, for example, with $\sin\alpha \simeq 0$ and $\tan\beta \simeq 2.3$, which gives $r(h VV) \simeq 0.9$, $r(h f \bar f) \simeq 1.1$, $r(h f_4 \bar f_4) \simeq 0$, which provides a cross section and decay widths similar to the SM Higgs at large.
It will take substantial amount of data and time to determine the Higgs production and branching ratio precisely enough to distinguish among the models containing SM-like Higgs.

Other Higgs bosons could be insensitive to the current LHC data depending on the mass and mixing with a Higgs singlet as extra decay modes such as decays into 4G fermion pairs or $Z Z'$, $Z' Z'$ can make considerable changes if kinematically accessible.
The latter can be dominant in a similar fashion that $H_\text{SM} \to W W$, $Z Z$ are dominant for a sufficiently heavy Higgs boson in the SM as the Goldstone boson equivalence theorem predicts.
Detailed study of the Higgs sector exploring various parameter choices will be a natural subject of subsequent studies.

Higgs decays into 4G fermions, if they are open, are also distinct in the presence of the 4G-parity.
For example, a decay channel could be $\higgs \to e_4^+ e_4^- \to W^+ \nu_4 W^- \bar \nu_4 \to 2 W$ with sizable missing transverse energy (\met).

DARK MATTER SECTOR:
The L4P, which we take as $\nu_4$, is a heavy neutral particle which is stable under the 4G-parity, and it can be considered as a DM candidate either by itself or as a fraction of what comprises about 23\% of the energy budget of the Universe.

Unfortunately, as has been well known from the early days \cite{Lee:1977ua}, the relic density of the thermal neutrino DM with $m_{\nu_4} > m_Z / 2$ is very small compared to the experimentally measured value due to the efficient annihilation via the $Z$ boson as well as the Higgs boson.
This holds even for very heavy $\nu_4$ since new channels such as $WW$ and $ZH$ open \cite{Enqvist:1988we}.
Thus, the $\nu_4$ can exist only as a subdominant DM in the standard cosmology.

Furthermore, the null results of the direct DM search experiments using the nuclear recoil \cite{Angle:2008we,Ahmed:2009zw,Aprile:2011hi} provide stringent bound on any massive neutrino DM candidate for both purely Dirac type and Majorana type.
A major direct search channel for the 4G neutrino DM candidate is the one mediated by the $Z$ boson in the $t$-channel.
This gives simply too large a cross section (for example, for a purely Dirac neutrino, $\sigma_{\nu-\text{nucleon}} \sim 0.1 G_F^2 m_\text{eff}^2 \sim 10^{-38} ~\text{cm}^2$ when we assume it is the sole DM candidate) to be consistent with the experimental bounds (roughly $10^{-44} ~\text{cm}^2$ level).
For a fermionic DM candidate with Dirac mass much larger than the Majorana mass, the nuclear recoil can be inelastic though, and it may escape the bounds from the direct detection experiments \cite{TuckerSmith:2001hy,Cui:2009xq}.
While the DM constraint is interesting and worth studying in detail, we will not pursue the details or try to fix the L4P mass in this paper.
For some recent studies on the heavy neutrino DM candidates, see Refs.~\cite{Kainulainen:2006wq,Belanger:2007dx,Zhou:2011fr,Lee:2011jk}.

PRODUCTIONS AND DECAYS OF 4G FERMIONS:
In analogy to the $R$-parity preserving supersymmetric models, the 4G fermion with 4G-parity can be produced only in pairs, and its decay chain always ends up with a stable L4P (see Fig.~\ref{fig:ProductionDecay}).
Decay modes depend on the mass spectrum, and we will assume $m_{\nu_4} < m_{e_4}$, $m_{d_4} < m_{u_4}$ in this paper for definiteness.
Some of the production (at hadron colliders) and decay modes are the following.

(1) Production:
$g g \to q_4 \bar q_4$, $q \bar q \to \gamma^*/Z^* \to e^+_4 e^-_4 / q_4 \bar q_4$, $u \bar d \to W^{+ *} \to \nu_4 e_4^+ / u_4 \bar d_4$.

(2) Decay:
$e_4^- \to W^- + \nu_4$, $u_4 \to W^+ + d_4$, $d_4 \to \bar u_a + \bar d_b + \bar \nu_{4} ~ (a, b = 1\sim3)$.

The light 4G quark $(d_4$) decay can occur through nonrenormalizable operators such as effective $QQQL$, $UUDE$, $UDDN$.
The $d_4$ decay depends on $x$.
With $x = -1/3$, it can decay through $\frac{1}{\Lambda^2} Q_aQ_bQ_4L_4$ and $\frac{1}{\Lambda^2} U_aD_bD_4N_4$.
Though the decay is through nonrenormalizable operators, it can decay instantaneously in the detector roughly for $\Lambda \lsim 100 ~\tev$, with $m_{d_4} - m_{\nu_4} \sim 100 ~\gev$.

It would be possible to construct nonrenormalizable operators with only the SM fermions that can affect proton decay if the universal cutoff scale is not very large.
However, it is also possible to realize the intermediate scale $\Lambda$ that is relevant only for the 4G involved processes (for example, with intermediate scale scalar fermions and appropriate coefficients in the supersymmetry framework) and thereby leaves the proton decay intact.

Since the $Z'$ does not couple to the SM fermions directly, the 4G fermions cannot form dilepton or other $Z'$ resonances \cite{Barger:2009xg,Barger:2011tz} through the Drell-Yan process unless we consider mixing effect.
The followings may be interesting channels to search for 4G fermions with the same mass hierarchy as before.
\bea
&&g g \to u_4 \bar u_4 \to W^+ W^- + 4 j ~(\text{2 $u$, $d$-type}) + \met ~~~~ \label{eq:mode1} \\
&&g g \to d_4 \bar d_4 \to 4 j ~(\text{2 $u$, $d$-type}) + \met ~~~~ \\
&&q \bar q \to \gamma^*/Z^* \to e_4^+ e_4^- \to W^+ W^- + \met ~~~~ \\
&&u \bar d \to W^{+ *} \to \nu_4 e_4^+ \to W^+ + \met ~~~~
\eea
Just like the supersymmetry search under the $R$-parity, sizable $\met$ accompany all the 4G search modes.

Because these channels are so different from the conventional 4G case, the 4G fermions could have escaped the experimental searches based on the SM4.
Furthermore, if the mass difference among the 4G fermions are sufficiently small, it may result only in very soft jets and off-shell $W$ bosons.
In hadron colliders such as the Tevatron and the LHC, such a soft jet is hard to distinguish from the backgrounds \cite{soft}.
Hence we may have quite light 4G fermions consistent with the existing experimental bounds.
Quantitative studies including comparison to the relevant background are called for to go beyond what we have discussed.

SUMMARY AND OUTLOOK:
In this paper, we introduced 4G-parity and its $U(1)$ gauge origin, which interacts with the 4G fermions but not with the SM fermions.
An additional Higgs doublet is required to be compatible with the recently measured $125-126 ~\gev$ Higgs signals, and the $U(1)$ charges of fermions are uniquely determined if the $U(1)$ charge of the additional Higgs doublet is fixed.

The 4G-parity allows a simple way to harbor (possibly light) 4G fermions while satisfying various theoretical and experimental constraints including the LEP $Z$ width bound.
Since the light extra fermions are allowed, it would be straightforward to extend the idea to include more than one extra generation of fermions.
The 4G-parity in the 4G scenario can be compared with the $R$-parity in the supersymmetry scenario.
The 4G particles are pair produced, and their decays end up yielding stable lightest 4G particles.
The massive 4G neutrino is taken as a very natural L4P, which would appear as the missing transverse energy in collider experiments.
Rich and distinct phenomenology is guaranteed and we briefly sketched some of that for the LHC experiments.
Among the open issues that require further consideration are the detailed study of the Higgs mixing effects, repercussions for CP violation and baryogenesis, detailed dark matter study, quantitative collider study, and ultraviolet completion of the model.

\begin{acknowledgements}
Acknowledgments:
We thank H. Davoudiasl for very helpful discussions.
HL thanks KIAS for the hospitality during the KIAS Phenomenology Workshop in 2011.
This work was supported in part by the United States DOE under Grant No.~DE-AC02-98CH10886, No.~DE-AC05-06OR23177 and by the NSF under Grant No.~PHY-1068008.
\end{acknowledgements}


\end{document}